\documentclass[aps,twocolumn,showpacs]{revtex4}
\usepackage{graphicx}   
\usepackage{latexsym}   

\newcommand{\beq}{\begin{equation}}
\newcommand{\eeq}{\end  {equation}}

\newcommand{\beqar}{\begin{eqnarray}}
\newcommand{\eeqar}{\end  {eqnarray}}

\newcommand{\SKIP}[1]{}		       		
\newcommand{\bld}[1]{\mbox{\boldmath $#1$}}	
\newcommand{\code}{{\tt FREYA}}			
\newcommand{\CGMF}{{\tt CGMF}}			
\newcommand{\FIFRELIN}{{\tt FIFRELIN}}		
\newcommand{\etal}{{\em et al.}}		

\begin{document}
~\

\title{Generation of fragment angular momentum in fission}

\author{J{\o}rgen Randrup$^1$ and Ramona Vogt$^{2,3}$}

\affiliation{
  $^1$Nuclear Science Division, Lawrence Berkeley National Laboratory,
  Berkeley, CA 94720, USA\break
  $^2$Nuclear \&\ Chemical Sciences Division,
  Lawrence Livermore National Laboratory, Livermore, CA 94551, USA\break
  $^3$Physics and Astronomy Department, University of California,
  Davis, CA 95616, USA}

\date{\today}

\begin{abstract}
A recent analysis of experimental data \cite{Nature} found that the
angular momenta of nuclear fission fragments are uncorrelated.
Based on this finding, the authors concluded that the spins
are therefore determined only {\em after} scission has occurred.
We show here that the nucleon-exchange mechanism,
as implemented in the well-established event-by-event fission model \code,
while agitating collective rotational modes in which the two
spins are highly correlated, 
nevertheless leads to fragment spins that are largely uncorrelated.
This fact invalidates the reasoning of those authors.
Furthermore, it was reported \cite{Nature} that the mass dependence 
of the average fragment spin has a sawtooth structure.
We demonstrate that such a behavior naturally emerges
when shell and deformation effects are included 
in the moments of inertia of the fragments at scission.
\end{abstract}

\maketitle

\section{Introduction}

A recent article by Wilson \etal\ \cite{Nature} 
addressed the generation of angular momentum in nuclear fission.
In particular, based on the analysis of unique and extensive experimental data 
taken at the ALTO facility of the IJC Laboratory in Orsay,
the authors concluded that there is no significant correlation 
between the angular momenta (or ``spins'') of the two fission fragments.
Because the authors assume that any spin generated prior to scission 
must result in strongly correlated fragment spins,
they conclude that their observation implies that the fragment spins
must be generated {\em after} the two fragments are no longer in contact.

Using the fission model \code\ \cite{PRC80},
we demonstrate here that a fission treatment based on correlated
rotational modes in the dinuclear complex {\em prior} to scission may in fact
endow the fragments with spins that are nevertheless approximately independent.
This fact invalidates the above assumption 
and hence casts doubt on the central conclusion in Ref.\ \cite{Nature}
regarding the mechanism for the angular momentum generation.

Furthermore, the data analysis in Ref.\ \cite{Nature} revealed that
the average magnitude of the fragment spins have a sawtooth-like dependence
on the mass number of the fragment.
We show that when realistic moments of inertia are employed
for the fledging fragments at scission,
the calculation naturally yields a sawtooth behavior of the mean spin.
This in turn provides an explanation for the observation in Ref.\ \cite{Nature}
that the light fragment may carry more angular mometum than its heavy partner.

In Sect.\ \ref{model} we recall
the main relevant features of the theoretical treatment.
Then, in Sect.\ \ref{corels},
we discuss the correlations between 
the fragment spin magnitudes as well as their directions.
In Sect.\ \ref{A-dep}, we address the mass-number dependence 
of the mean spin magnitude
and we finally present a concluding discussion in Sect.\ \ref{disc}.

\section{Model}
\label{model}

The fission model \code\ \cite{PRC80,FREYA2} employs Monte Carlo techniques 
to the selection of the mass, charge, and velocity of the primary fragments
as well as to their subsequent decays 
by sequential neutron evaporation and photon radiation,
thereby generating large samples of complete fission events.
The model is in many respects similar to other fission simulation treatments,
such as \CGMF\ \cite{CGMF}, \FIFRELIN\ \cite{FIFRELIN}, and GEF\ \cite{GEF},
although these model differ in many details.
For example, \code\ is the only one that conserves angular momentum
thoughout an event.

Of particular relevance is the selection of the fragment spins \cite{PRC89},
which was recently discussed in some detail \cite{PRC103}.
The procedure employed in \code\ 
is motivated by the Nucleon-Exchange Transport model \cite{NPA327,NPA383}
which considers the effect of multiple nucleon transfers
between the two parts of a dinuclear complex.
Because each nucleon carries baryon number (and possibly electric charge)
as well as linear and angular momentum,
the associated observables exhibit a diffusive evolution
as a function of the number of transfers.
This mechanism was found to be the dominant cause of dissipation
in strongly damped nuclear reactions \cite{Huizenga-1984}
and it is expected to also play an important role 
during the late stages of fission,
as the system develops a binary character prior to scission \cite{AP113}.

A detailed study of the effect of nucleon exchange
on the dynamical evolution of the fragment spins 
considered the agitation of the six dinuclear rotational modes
(the transverse modes wriggling and bending, which are doubly degenerate,
and the coaxial modes twisting and tilting) \cite{NPA433I,NPA433II}.
In particular, expressions were derived 
for the associated mobility coefficients which determine the time scales.
When the relaxation time for a particular mode is short 
compared with the collective evolution,
the associated spin distribution quickly readjusts to the evolving geometry
and so it remains in local equilibrium.

Invoking this idealized limit for all six dinuclear modes,
Moretto and Schmitt \cite{MorettoPRC21}
formulated a statistical equilibrium model 
for the fragment spins in fission and heavy-ion reactions.
However, the mobility coefficients for the dinuclear rotational modes
differ significantly in magnitude,
causing the relaxation time for the wriggling mode to be very fast,
whereas the tilting mode is being agitated only quite slowly.
As a rough way of taking this complexity into account,
\code\ implements the effect of the nucleon-exhange mechanism
by assuming full relaxation of the transverse modes
(wriggling and bending), while the coaxial modes (twisting and tilting)
are not agitated \cite{PRC87,PRC89}.

Generally, the fissioning complex has an overall angular momentum $\bld{S}_0$,
of which the fragments receive their share,
$\underline{\bld{S}}_f=({\cal I}_f/{\cal I}_{\rm tot})\bld{S}_0$,
where ${\cal I}_f$ is the moment of inertia of the fragment at scission
and 
${\cal I}_{\rm tot}={\cal I}_L+{\cal I}_H+{\cal I}_R$
is the total moment of inertia,
with ${\cal I}_R$ being the moment of inertia for the relative fragment motion.\
These rigid-rotation contributions tend to be negligible
relative to the contributions from the statistical fluctuations.

The sampling of those is complicated by the fact that
the individual fragment spins are not independent, 
due to the coupling imposed by conservation.
There are three angular momentum vectors at scission:
the spins of the two dinuclear partners, $\bld{S}_L$ and $\bld{S}_H$,
and their relative angular momentum, $\bld{L}$.
Because the system is isolated, their sum is conserved,
$\bld{S}_L+\bld{S}_H+\bld{L}=\bld{S}_0$,
leaving then six rotational degrees of freedom.
These are conveniently represented by the six dinuclear normal modes
in terms of which the rotational part of the collective Hamiltonian is diagonal.

The modes considered
(all six modes in the statistical treatment \cite{MorettoPRC21}
and only the four transverse modes in \code) are then populated statistically.
Thus the amplitude $s_m$ of a given mode $m$ is sampled from
$P_m(s_m)\sim\exp(-s_m^2/2{\cal I}_mT_S)$, 
where ${\cal I}_m$ is the moment of inertia for that mode
\cite{MorettoPRC21,PRC89}.
The wriggling moment of inertia is ${\cal I}_+
=({\cal I}_L+{\cal I}_H)({\cal I}_L+{\cal I}_H+{\cal I}_R)/{\cal I}_R$,
while that for bending is ${\cal I}_-
={\cal I}_L{\cal I}_H/({\cal I}_L+{\cal I}_H)$.
(The temperature $T_S$ employed by \code\
takes account of the fact that the distortion of the fragments at scission 
reduces the available statistical energy somewhat \cite{PRC87,PRC89}.)
Once the normal modes have been agitated,
yielding $\bld{s}_+$ for wriggling and $\bld{s}_-$ for bending,
the individual fragment spins then readily follow \cite{PRC89},
resulting in
\beq
\bld{S}_f = \underline{\bld{S}}_f
+({\cal I}_f/{\cal I}_{\rm tot})\bld{s}_+ \pm \bld{s}_-\ ,
\eeq
and their variances are
$\sigma_f^2=
2(1-{\cal I}_f/{\cal I}_{\rm tot}){\cal I}_fT_S$.

\section{Spin correlations}
\label{corels}

\begin{table}
\begin{tabular}{c|crrrrc}\hline\hline\\[-2ex]
{\rm Case:} &~&
$^{235}$U($n$,f) & $^{238}$U($n$,f) & $^{239}$Pu($n$,f) & $^{252}$Cf(sf) &\\[1ex]
\hline\\[-2.5ex]
$\overline{S}_L=\langle S_L\rangle$ & 
	   &  4.27~  &  4.43~  &  4,58~  &  5.08~ &\\[-0.8ex]
&	   & (6.08) & (6.59) & (6.86) & (7.48)\\
$\overline{S}_H=\langle S_H\rangle$ & 
	   &  5.66~  &  5.80~  &  5.93~  &  6.33~ & \\[-0.8ex]
&	   & (5.31) & (5.49) & (5.60) & (6.80)\\
$c(S_L,S_H)$ (\%) &
	   &  0.2~  &  0.2~ &  0.1~  &  0.1~ &\\[-0.8ex]
&	   & (-10.9) & (-10.8) & (-10.3) & (0.8)\\
$f_1$ (\%) &
	   &  -8.2~  & -8.3~ & -8.3~  &  -8.4~ &\\[-0.8ex]
&	   & (-10.9) & (-11.3) & (-11.7) & (-13.5)\\[1ex]
\hline\hline
\end{tabular}
\caption{The average magnitudes of the primary fission fragment spins,
$\overline{S}_L$ and $\overline{S}_H$,
and the associated correlation coefficients
$c(\overline{S}_L,\overline{S}_H)$ for four fission cases:
$^{235}$U($n$,f), $^{238}$U($n$,f),$^{239}$Pu($n$,f), and $^{252}$Cf(sf), as
obtained with \code\ using either monotonically increasing moments of inertia 
or moments of inertia with a refined $A$ dependence (numbers in parantheses).
Also shown are the amplitudes $f_1$ in the distribution of the
spin opening angles, $P(\phi_{LH})=1+f_1\cos\phi_{LH}$.
}\label{t:S}
\end{table}

\begin{figure}[tbh]	   
\includegraphics[width=0.95\columnwidth, trim = {0.1cm, 0cm, 0.1cm, 0},clip]{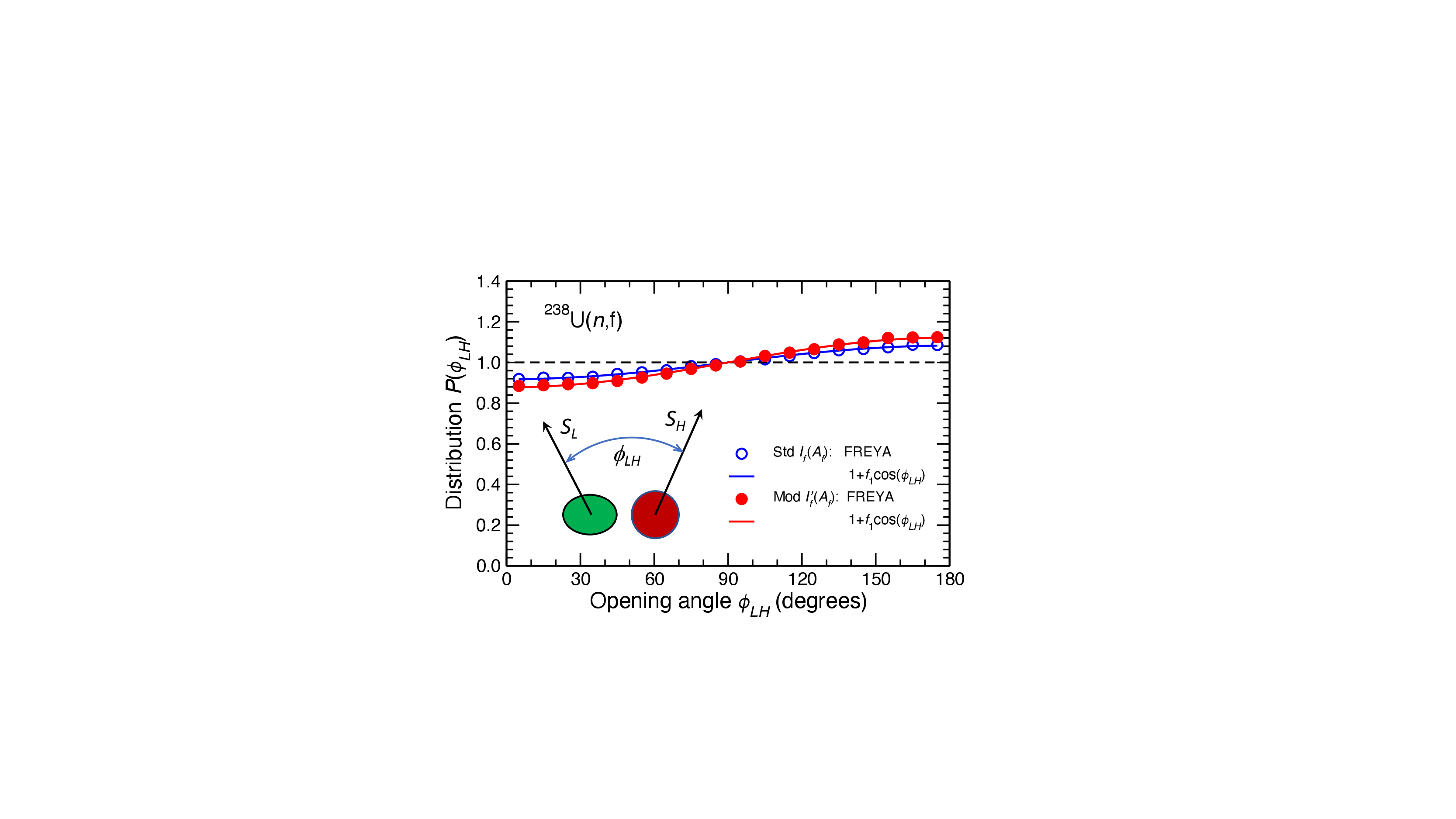}
\caption{(Color online)
The distribution of the opening angle $\phi_{LH}$ 
between the angular momenta of the two fission fragments 
from $^{238}$U($n$,f), as obtained by \code\ using 
either the standard moment of inertia, 
${\cal I}_f(A_f)=0.5\,{\cal I}_{\rm rigid}(A_f)$ (dots)
or the modified form depending on the shapes of the fragments at scission,
${\cal I}'_f(A_f)$ (open circles).
Also shown are the lowest-order Fourier approximations, 
$P(\phi_{LH}) = 1 + f_1\cos\phi_{LH}$.
}\label{f:phiLH}
\end{figure}

As discussed above, the two fragment spins are not independent,
due to angular momentum conservation.
Indeed, the spin contributions from the two fragments to wriggling
are perfectly parallel,
while the contributions to bending are anti-parallel.
When these dinuclear modes are populated statistically,
the resulting correlation coefficient 
for the individual fragment spins amounts to
\beqar
c(\bld{S}_L,\bld{S}_H) \!&\equiv&\!
[\langle\bld{S}_L\!\cdot\!\bld{S}_H\rangle
- \langle\bld{S}_L\rangle\langle\bld{S}_H\rangle]/[\sigma_L\sigma_H]\\ \nonumber
\!&=&\! -\left\{{\cal I}_L{\cal I}_H/
	[({\cal I}_R+{\cal I}_L)
	 ({\cal I}_R+{\cal I}_H]\right\}^{1/2}\! .
\eeqar
This quantity is generally rather small 
because the moment of inertia for the relative
fragment motion, ${\cal I}_R$, is typically an order of magnitude larger than
those of the individual fragments, ${\cal I}_L$ and ${\cal I}_H$,
${\cal I}_R\gg{\cal I}_f$.
Thus, even though the fragment spins are strongly coupled for each of the
normal modes, the resulting spins are expected to be relatively independent.
This expectation is indeed borne out by actual \code\ simulations
\cite{PRC103},
as illustrated in Table \ref{t:S} and Fig.\ \ref{f:phiLH}.

The magnitudes of the fragment spins, $S_f=|\bld{S}_f|$,
have rather wide distributions with average values 
$\overline{S}_f\equiv\langle S_f\rangle\approx5-6\,\hbar$.
The associated covariance is given by 
$\sigma(S_L,S_H)=
\langle S_LS_H\rangle-\overline{S}_L\overline{S}_H$,
so the spin magnitude correlation coefficient is
$c(S_L,S_H)=
\sigma(S_L,S_H)/[\sigma(S_L)\sigma(S_H)]$.
Table \ref{t:S} lists the average spin magnitudes and 
the correlation coeficient for four fission cases of frequent interest.
The correlation coefficients are essentially zero,
indicating that the primary spin magnitudes are largely uncorrelated,
in accordance with the reported experimental finding \cite{Nature}.
(Unfortunately, the mutual independence of the two fragment spins
is not quantified in Ref.\ \cite{Nature}, so it is not possible to
make a precise comparison.) 

While the experimental data \cite{Nature} cannot yield information
on the fragment spin directions, 
it is noteworthy that those are also largely uncorrelated in \code.
The degree of correlation between the fragment spin directions 
is brought out by the distribution of the opening angle 
between the two fragment spins, $\phi_{LH}$, 
which is given by $\cos\phi_{LH}
=\bld{S}_L\!\cdot\!\bld{S}_H/[S_LS_H]$.
This function is shown in Fig.\ \ref{f:phiLH} for $^{238}$U($n$,f).
(The other cases look very similar.)
As was shown recently \cite{PRC103},
the undulation of the fragment spin opening angle
is generally well represented by the first harmonic,
$P(\phi_{LH})\approx1+f_1\cos\phi_{LH}$.
Table \ref{t:S} shows the amplitudes $f_1$ for the four cases considered.
As seen, they are all rather small, being of the order of 10\%,
showing that the spin directions are also largely uncorrelated.

\section{Mass dependence}
\label{A-dep}

\begin{figure}[tbh]
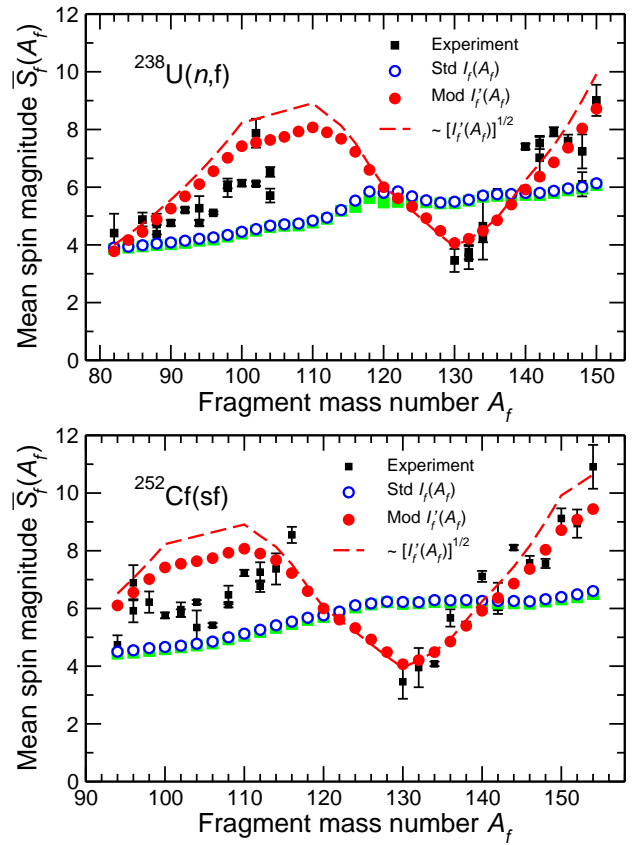
	   
\includegraphics[width=0.95\columnwidth]{SofA-U239.eps}
\includegraphics[width=0.95\columnwidth]{SofA-Cf252.eps}
\caption{(Color online) 
The average magnitude of the spin of a fragment
as a function of its mass number, $\overline{S_f}(A_f)$,
for $^{238}$U($n$,f) and $^{252}$Cf(sf).
The solid squares are the data \cite{Nature}.
The \code\ calculations using the standard moments of inertia, 
${\cal I}_f(A_f)=0.5\,{\cal I}_{\rm rigid}(A_f)$
are shown by the open circles, with the green squares showing
the values after the completion of the neutron evaporation.
In addition to the spin values (dots) 
calculated with the modified moments of inertia,
${\cal I}'_f(A_f)$, 
the dashes show the (rescaled) function $[{\cal I}'_f(A_f)]^{1/2}$.
}\label{f:SofA}
\end{figure}

In addition to the qualitative finding that 
the fragment spins are largely uncorrelated,
Ref.\ \cite{Nature} presented data
for the dependence of the average spin magnitude
$\overline S_f$ on the fragment mass number $A_f$, 
revealing a sawtooth-like structure of $\overline{S}_f(A_f)$
(see Fig.\ \ref{f:SofA}).
We now show that such a behavior emerges naturally
due to the variation  of the moments of inertia of the fragments at scission.
\SKIP{
Fragments near the doubly magic nucleus $^{132}$Sn 
have spherical equilibrium shapes with a high degree of rigidity
and abnormally low level densities, 
so they do not become very distorted at scission
and cannot carry much angular momentum.
As one moves away from this special region,
the fragments tend to have ever larger equilibrium deformations and,
moreover, they are softer and become further distorted at scission,
so they can accommodate significantly more angular momentum.
(This qualitative expectation was indeed borne out by recent microscopic 
studies of scission comnfigurations \cite{AlbertssonPRC103,Bulgac,Marevic}.)
}

When the dinuclear rotational modes are sampled from thermal distributions,
as in \code, then
$\langle S_f^2\rangle\approx2{\cal I}_f T_S$ (see above),
so the mean fragment spin scales approximately 
as the square root of the moment of inertia, ${\cal I}_f$.
To illustrate the close connection between 
$\overline{S}_f$ and ${\cal I}_f$,
we first note that the standard version of \code\ \cite{FREYA2}
assumes that the moment of inertia of a fragment is given by 
${\cal I}_f(A_f)=c_{\rm rot}{\cal I}_{\rm rigid}(A_f)$,
with the global reduction factor being $c_{\rm rot}=0.5$.
Thus the $A_f$ dependence of the moment of inertia is monotonic and
the calculated $\overline{S}_f(A_f)$ is rather featureless.
This is illustrated in Fig.\ \ref{f:SofA} where 
$\overline{S}_f(A_f)$ obtained with the standard version of \code\ 
is shown for $^{238}$U($n$,f) and $^{252}$Cf(sf) 
together with the data reported in Ref.\ \cite{Nature}.
[The figure also shows that the effect of neutron evaporation 
on the spin magnitudes is negligible, 
thus lending support to the assumption made in the data analysis.]

However, as pointed out in Ref.\ \cite{Nature},
fragments near the doubly magic nucleus $^{132}$Sn
have rather rigid spherical shapes and abnormally low level densities,
as a consequnece of which they cannot accomodate very much angular momentum.
Indeed, in order to match the measured $\overline{S}_f$ values 
near $^{132}$Sn,
it is necessary to reduce the rigid moment of inertia by 0.2 rather than 0.5.
On the other hand, away from this special region the fragments
have deformed equilibrium shapes 
and are likely significantly further distorted at scission.
Thus they have considerably larger moments of inertia and
can acquire correspondingly larger angular momenta.
(These expectations were indeed borne out by recent microscopic studies 
of scission comnfigurations \cite{AlbertssonPRC103,Bulgac,Marevic}.)

For illustrative purposes, we have roughly incorporated these effects
into modified moments of inertia, ${\cal I}_f'(A_f)$,
using microscopic results on the fragment equilibrium shapes and,
as in the standard \code, one global scaling parameter.
The resulting mean spin magnitudes then acquire a sawtooth-like behavior
that roughly matches the experimental data,
as illustrated in Fig.\ \ref{f:SofA}.

Also shown is $[{\cal I}_f'(A_f)]^{1/2}$
(rescaled for convenience), bringing out the close connection between
the moment of inertia and the mean spin magnitude.
The remarkably similar mass dependence of those two quantities suggests
that the measurements of $\overline{S}_f(A_f)$ in effect
provide information on the geometry of the scission configurations.

Finally it should be noted that the modified moments of inertia
tend to favor the light fragments (because of their deformations),
so these will now have larger spins than their heavy partners,
as found experimentally \cite{Nature}
(This is also consistent with recent microscopic studies \cite{Bulgac,Marevic}.)
However, importantly, fragment spins remain approximately uncorrelated
after the modification of the moments of inertia,
as illustrated in Table \ref{t:S} and Fig.\ \ref{f:phiLH}.
\SKIP{
(The correlation coefficient $c(\overline{S}_L,\overline{S}_H)$
changes from being below one per cent to around ten per cent
and the directional correlation amplitude $f_1$
changes from $\approx-8.5\%$ to $\approx-11\%$.)
}

\section{Discussion}
\label{disc}

The mean-field character of low-energy nuclear collective motion
implies that the associated dissipation is of one-body form \cite{AP113}.
For a single nucleus, the one-body dissipation results from
particle-hole excitations generated by the evolving mean field.
For binary systems, such as those encountered in nuclear reactions and fission,
an additional one-body mechanism 
is the multiple transfer of individual nucleons 
between the two parts,
a process that leads to a diffusive evolution of
the mass and charge partition in the dinucleus
as well as the linear and angular momenta of the two partners.
Because the active nucleons reside in the Fermi surface
they carry relatively large momenta and, as a consequence,
the resulting dissipation is strong \cite{AP113,PRL44}.

This general character of nuclear dynamics causes the shape evolution 
in fission to be rather slow,
a feature \code\ seeks to incorporate 
by assuming that the transverse dinuclear rotational modes
maintain full equilibrium up to scission,
while the coaxial modes are insignificantly agitated.
It is important to note that even though the contributions from each 
dinuclear normal mode to the individual fragment spins are strongly aligned
the resulting fragment spins are largely uncorrelated.
This is in accordance with the experimental finding in Ref.\ \cite{Nature}.

This finding demonstrates that the absence of spin correlation 
can {\em not} be taken as evidence for
the fragments having acquired their spins independently
and it thus invalidates the assumption underlying 
the data interpretation advanced in Ref.\ \cite{Nature}.
(In fact, 
the independent population of rotational states in the fragments
{\em after} their separation 
is hard to reconcile
with the principle of angular momentum conservation for isolated systems.)

We have furthermore demonstrated that the sawtooth-like mass dependence
of the average fragment spin can be understood 
as a reflection of the moments of inertia 
of the fragments at the time of their formation.
Such a behavior would arise also in other simulation codes
(such as Refs.\ \cite{CGMF,FIFRELIN,GEF})
if similar moments of inertia were were employed.
(It should also be noted that $\overline{S}_f(A_f)$ would still
display a sawtooth form even in the extreme scenario
where only the wriggling modes were populated,
causing the two fragment spins to be perfectly parallel in each event.)
The finding, based on the observed sawtooth behavior of $\overline{S}_f(A_f)$,
that the light fragment carries more angular momentum than its heavy partner,
was anticipated recently on the basis of microscopic studies
\cite{Bulgac,Marevic}.

Finally we have pointed out that
the strong connection between the moments of inertia of the fledging fragments
and their mean spin magnitudes suggest that measurements of 
$\overline{S}_f(A_f)$, 
such as those reported in Ref.\ \cite{Nature},
may provide unique experimental information on 
the fissioning system at the time of scission,
which in turn would be very helpful 
for the further development of fission theory.

~\\
\noindent{\bf Acknowledgments}\\
This work was supported by the Office of Nuclear Physics 
in the U.S.\ Department of Energy under Contracts DE-AC02-05CH11231 (J.R.) 
and DE-AC52-07NA27344 (R.V.) and was supported by 
the DOE Office of Nuclear Nonproliferation and the LLNL LDRD Program 
under Project No.\ 20-ERD-031 (R.V.).



			\end{document}